\documentclass[12pt]{article}
\usepackage{amssymb,amsmath}
\usepackage{color}

\numberwithin{equation}{section}
\newtheorem{theorem}{Theorem}[section]     
\newtheorem{definition}[theorem]{Definition}

\newtheorem{proposition}[theorem]{Proposition}
\newtheorem{lemma}[theorem]{Lemma}

\newtheorem{corollary}[theorem]{Corollary}

\newtheorem{remark}[theorem]{Remark}

\def\sa{\sum_\alpha\epsilon_\alpha}

\def\wa{W_{\alpha}}
\def\wb{W_{\beta}}
\def\d{\partial}
\def\n{\noindent}
\def\f{\frac}

\def\na{\nabla}
\def\proof{\noindent\hspace{2em}{\itshape Proof: }}

\newcommand{\eqa}{\begin{eqnarray}}
\newcommand{\eeqa}{\end{eqnarray}}
\newcommand{\beq}{\begin{equation}}
\newcommand{\eeq}{\end{equation}}

\newcommand{\de}{\partial}

\begin{document}
\title{Purely non-local Hamiltonian formalism, \\Kohno connections and $\vee$-systems}
\author{Alessandro Arsie* and Paolo Lorenzoni**\\
\\
{\small *Department of Mathematics and Statistics}\\
{\small University of Toledo,}
{\small 2801 W. Bancroft St., 43606 Toledo, OH, USA}\\
{\small **Dipartimento di Matematica e Applicazioni}\\
{\small Universit\`a di Milano-Bicocca,}
{\small Via Roberto Cozzi 53, I-20125 Milano, Italy}\\
{\small *alessandro.arsie@utoledo.edu,  **paolo.lorenzoni@unimib.it}}

\date{}

\maketitle

\begin{abstract}
In this paper, we extend purely non-local Hamiltonian formalism to a class of Riemannian F-manifolds, without assumptions on the semisimplicity of the product $\circ$ or on the flatness of the connection $\nabla$. In the flat case we show that the recurrence relations for the principal hierarchy can be re-interpreted using a local and purely non-local Hamiltonian operators and in this case they split into two Lenard-Magri chains, one involving the even terms, the other involving the odd terms. 
Furthermore, we give an elementary proof that the Kohno property and the $\vee$-system condition are equivalent under suitable conditions and we show how to associate a purely non-local Hamiltonian structure to any $\vee$-system, including degenerate ones. 
\end{abstract}

\section{Introduction}

The study of Frobenius manifolds is an important branch of modern mathematics, with relations with many areas, ranging from singularity theory and Coxeter groups  to Gromov-Witten invariants and hierarchies of integrable PDEs (see \cite{du93}). 

Frobenius manifolds are given by the data $(M,\eta,\circ,e,E)$, where $M$ is a smooth manifold, $\eta$ a nondegenerate metric, $\circ$ is a smooth commutative associative product on sections of $TM$,  $e$ is the unit vector field of $\circ$ and $E$ is the Euler vector field. These structures are required to  satisfy some further compatibility axioms.  

In particular, in the theory of integrable systems  
 Frobenius manifolds describe the dispersionless limit of many important integrable hierarchies, called \emph{principal hierarchies}, having the form
 \begin{equation}\label{one}
 u_t=X\circ u_x
 \end{equation}
where $u=(u^1,\dots,u^n)$ and $X$ is a suitable vector field on $M$. 

The definition of the flows \eqref{one} does not rely on the full set of the axioms  entering the definition of  Frobenius manifold. 
Essentially what one needs is a symmetric connection $\nabla$ and a commutative associative product $\circ$ satisfying the conditions (see \cite{LPR})
\begin{eqnarray*}
\nabla_i c^k_{lj}&=&\nabla_l c^k_{ij},\qquad\forall i,j,k,l,\\
R^k_{lmi}c^n_{pk}+R^k_{lip}c^n_{mk}+R^k_{lpm}c^n_{ik}&=&0,\qquad\forall i,l,m,n,p,
\end{eqnarray*}
where $c^i_{jl}$ are the structure constants for the product $\circ$ and $R^k_{lmi}$ are the components of the Riemann tensor associated to the symmetric connection $\nabla$. 
In practice, in the most interesting examples, the connection $\nabla$ is flat, so the second condition above is automatically satisfied. 

In the case of a flat connection, the vector fields $X_{(\alpha,p)}$ defining the flows \eqref{one} are the  $\lambda$-coefficients of the expansion at $\lambda=\infty$
 of a basis of parallel vector fields
$$X_{(\alpha)}=\sum_k \f{X_{(\alpha,p)}}{\lambda^k}$$
of the deformed flat connection
\beq\label{dfc}
\tilde\nabla=\nabla+\lambda\circ.
\eeq

In flat coordinates for $\nabla$, the equations of the principal hierarchy take the form of a system of conservation laws
\begin{equation}
u^i_{t_{(\alpha,p)}}=\d_x X^i_{(\alpha,p+1)}.
\end{equation}

In order to define a Frobenius manifold one needs some additional axioms which
 are related to further properties of the hierarchy. Indeed
 \begin{itemize}
 \item the existence of a metric $\eta$ compatible with $\nabla$ and invariant with respect to $\circ$ is equivalent to the closure 
 of the $1$-forms $\eta_{ij}X^j$ \cite{AL-Hamiltonian}. Locally in this case we have
$$\eta_{ij}X^j=\d_i H$$
and thus
\begin{equation}
u^i_t=\eta^{ij}\d_x\d_j H.
\end{equation}
This means that all the flows are Hamiltonian w.r.t. the local Hamiltonian operator $\eta^{ij}\d_x$ defined
 by the flat metric $\eta$. 
\item the presence of a second contravariant flat metric $g$ (usually called \emph{intersection} form) is related to 
 the existence of a second compatible local Hamiltonian structure. 
Some authors distinguish between ``conformal'' Frobenius manifolds (with bi-Hamiltonian principal hierarchy)
and ``non-conformal'' Frobenius manifolds (with Hamiltonian principal hierarchy) not requiring for the latter the existence of the Euler vector field. 
In the conformal case, even if all the flows of the principal hierarchy are bi-Hamiltonian, in general they are not related by
bi-Hamiltonian recursive relations.
\end{itemize}
In both cases, besides the local Hamiltonian structure(s),
 there is anyway a purely non-local compatible Hamiltonian structure defined by (see \cite{mokhov}): 
\beq\label{PNLHSintro} 
P^{ij}=\eta^{lm}(u_x\circ)^i_l
\left(\frac{d}{dx}\right)^{\!-1}\!\!\!(u_x\circ)^j_m.
\eeq
In the first part of the paper we show that purely non-local Hamiltonian operators
$$P^{ij}=\sum_{\alpha=1}^n\epsilon_{\alpha}(X_{(\alpha)}\circ u_x)^i
\left(\frac{d}{dx}\right)^{\!-1}\!\!\!(X_{(\alpha)}\circ u_x)^j$$
 are related to a metric $\eta$ of the form:
$$\eta=\sum_{\alpha}\epsilon_{\alpha}X_{\alpha}\otimes X_{\alpha},$$
where $\epsilon_{\alpha}$ are $\pm1$ and the vector fields $X_{\alpha}$ are solutions of the equation 
$$c^i_{jk}\nabla_k X_{\alpha}^l=c^i_{kl}\nabla_j X_{\alpha}^l.$$
This generalizes a results obtained in \cite{GLR} in the semisimple case. In the flat case, $X_{\alpha}$ are flat vector fields and 
and thus the corresponding purely non-local Hamiltonian operator in flat coordinates assumes the form \eqref{PNLHSintro}.

Furthermore, we show that the ``even'' and the ``odd'' flows of the principal hierarchy give rise separately to recurrence chains with respect to the bi-Hamiltonian structure defined in flat coordinates $(u^1,\dots,u^n)$ by the local Hamiltonian operator $P_1=\eta^{ij}\d_x$ and by
the purely non-local Hamiltonian operator $P_2$ defined in \eqref{PNLHSintro}:
\begin{displaymath}
\begin{array}{rcl}
&&0=P\delta H_{(0,\alpha)}\\
&\stackrel{P_1}{\nearrow}&\\
\delta H_{(0,\alpha)}&&\\
& \stackrel{P_2}{\searrow} &\\
&& P_2\delta H_{(0,\alpha)}=P_1 \delta H_{(2,\alpha)} \\
&\stackrel{P_1}{\nearrow}&\\
\delta H_{(2,\alpha)}&&\\
& \stackrel{P_2}{\searrow} &\\
&& P_2 \delta H_{(2,\alpha)}=P_1\delta H_{(4,\alpha)} \\
&\stackrel{P_1}{\nearrow}&\\
.\,.\,.&&
\end{array}
\qquad\qquad
\begin{array}{rcl}
\delta H_{(1,\alpha)}&&\\
& \stackrel{P_2}{\searrow} &\\
&& P_2 \delta H_{(1,\alpha)}=P_1 \delta H_{(3,\alpha)} \\
&\stackrel{P_1}{\nearrow} & \\
\delta H_{(3,\alpha)}&& \\
&\stackrel{P_2}{\searrow} & \\
&& P_2\delta H_{(3,\alpha)}=P_1 \delta H_{(5,\alpha)} \\
&\stackrel{P_1}{\nearrow} & \\
\delta H_{(5,\alpha)}&& \\
&\stackrel{P_2}{\searrow} & \\
&&  .\,.\,.
\end{array}
\end{displaymath}

In the second part of the paper, we briefly discuss
 the notion of $\vee$-systems introduced by Veselov in \cite{Ve, Ve2} to construct solutions of the
generalized WDVV equations and  the so called Kohno connection.

We prove that there is a one to one correspondence between $\vee$-systems $\mathcal{V}$ on one hand and suitable classes of data satisfying the Kohno condition on the other, thus clarifying the relationship between these two constructions (see Theorem \ref{veeKohno}). Therefore, the $\vee$-condition can be traded for the Kohno condition as long as the set of covectors forming the $\vee$-systems are a spanning set for the dual vector space. 

The Kohno connection associated to a $\vee$-system is a special case of deformed flat connection \eqref{dfc}. Therefore, starting from Kohno connection
 one can immediately reconstruct  all the geometric data defining a non-conformal Frobenius manifold.

In this case, calling $\alpha$ the covectors spanning the $\vee$-system $\mathcal{V}$, we have that the (contravariant) metric is given by the following quadratic expansion
$$\check{G}=\sum_{\check{\alpha}\in\mathcal{V}}\check{\alpha}\otimes\check{\alpha},$$
while the product $\circ$ is defined by the structure constants
$$c^i_{lp}(u)=\sum_{\alpha\in \mathcal{V}}\frac{\alpha_l \, \alpha_p \, \check{\alpha}^i}{\alpha(u)},$$
where  $\alpha_i=\check{G}^{-1}_{ij}\check{\alpha}^j$.
 
Furthermore, applying the results of the first part of the paper, we prove that there is a purely non-local Hamiltonian operator $P^{\mathcal{V}}$ associated to any $\vee$-system $\mathcal{V}$ having the form: \beq\label{PNHSveeintro}
 P^{\mathcal{V}}=\sum_{\beta, \gamma\in \mathcal{V}}\check{G}(\beta, \gamma) \partial_x \log(\beta(u)) \partial_x^{-1} (\partial_x \log(\gamma(u))) \check{\beta} \otimes \check{\gamma}.
\eeq

We also analyze two degenerate cases of $\vee$-systems corresponding to Lie superalgebras and we show that also to the degenerate cases we can associate a non-conformal Frobenius structure without unity and consequently a purely non-local Hamiltonian operator of the form \eqref{PNHSveeintro}.

\section{Purely non-local Hamiltonian formalism}

Purely non-local Hamiltonian formalism was studied in \cite{mokhov,GLR}.


Let 
$W_{\alpha},\alpha=1,...,N$ be non-degenerate 
$(1,1)$-tensors, then the  bivector
\beq\label{NLPHT}
P^{ij}=\sum_\alpha\epsilon_\alpha 
\,W_\alpha(u)^i_s\,u^s_x\,\,\d_x^{-1}\,\,W_\alpha(u)^j_l 
\,u^l_x,
\eeq
is a Poisson bivector if the following  conditions are satisfied (see \cite{GLR}):
\begin{eqnarray}
\label{symmetry}
(W_\beta)^m_q\de_m(W_\alpha)^k_l+(W_\beta)^m_l\de_m(W_\alpha)^k_q+(W_\alpha)^k_m\de_l(W_\beta)^m_q+(W_\alpha)^k_m\de_q(W_\beta)^m_l \\
\nonumber=(W_\alpha)^m_q\de_m(W_\beta)^k_l+(W_\alpha)^m_l\de_m(W_\beta)^k_q+(W_\beta)^k_m\de_l(W_\alpha)^m_q+(W_\beta)^k_m\de_q(W_\alpha)^m_l,\\
\label{commutativity}
\left[W_\alpha,W_\beta\right]=0, \qquad \forall\, 
\alpha, \beta,\\
\label{zerocurv}
\sa\Big((W_\alpha)^i_k(W_\alpha)^j_h-(W_\alpha)^i_h(W_\alpha)^j_k\Big)=0.
\end{eqnarray}

To prove the main result of this section, we need the following 
\begin{lemma}\label{lemma1}
If the affinors $W_{\alpha}$ satisfy conditions \eqref{commutativity}, then the condition \eqref{symmetry} is tensorial, i.e. it is equivalent to 
$$(W_\beta)^m_q\na_m(W_\alpha)^k_l+(W_\beta)^m_l\na_m(W_\alpha)^k_q+(W_\alpha)^k_m\na_l(W_\beta)^m_q+(W_\alpha)^k_m\na_q(W_\beta)^m_l =
$$
$$=(W_\alpha)^m_q\na_m(W_\beta)^k_l+(W_\alpha)^m_l\na_m(W_\beta)^k_q+(W_\beta)^k_m\na_l(W_\alpha)^m_q+(W_\beta)^k_m\na_q(W_\alpha)^m_l,$$ where $\nabla$ is any symmetric connection.
\end{lemma}
\proof 
Expanding the covariant derivatives in terms of the Christoffel symbols, the condition \eqref{symmetry} is tensorial iff 
$$(\wb)^m_q\left(\Gamma^k_{mp}(\wa)^p_l-\Gamma^p_{ml}(\wa)^k_p \right)+(\wb)^m_l\left( \Gamma^k_{mp}(\wa)^p_q-\Gamma^p_{mq}(\wa)^k_p \right)+$$
$$+(\wa)^k_m\left(\Gamma^m_{lp}(\wb)^p_q-\Gamma^p_{lq}(\wb)^m_p \right)+(\wa)^k_m\left(\Gamma^m_{qp}(\wb)^p_l-\Gamma^p_{ql}(\wb)^m_p \right)=$$
$$=(\wa)^m_q\left(\Gamma^k_{mp}(\wb)^p_l-\Gamma^p_{ml}(\wb)^k_p \right)+(\wa)^m_l\left(\Gamma^k_{mp}(\wb)^p_q-\Gamma^p_{mq}(\wb)^k_p \right)+$$
$$+(\wb)^k_m\left(\Gamma^m_{lp}(\wa)^p_q-\Gamma^p_{lq}(\wa)^m_p \right)+(\wb)^k_m\left(\Gamma^m_{qp}(\wa)^p_l-\Gamma^p_{ql}(\wa)^m_p \right).$$
Using the symmetry of the Christoffel symbols with respect to lower indices everything cancels out except for the residual term
$$(\wa)^k_m\left(-\Gamma^p_{lq}(\wb)^m_p- \Gamma^p_{ql}(\wb)^m_p\right)=(\wb)^k_m\left(-\Gamma^p_{lq}(\wa)^m_p  -\Gamma^p_{ql}(\wa)^m_p\right).$$
But this equality is equivalent to $2\Gamma^p_{lq}[\wa, \wb]^k_p$ and so it holds provided that the conditions \eqref{commutativity} are fulfilled.

\begin{flushright}
$\Box$
\end{flushright}

Suppose now that the affinors $W_{\alpha}$ have the form
$$(W_{\alpha})^i_j=(X_{\alpha}\circ)^i_j=c^i_{jk}X_{\alpha}^k.$$
where $c^i_{jk}$ are the structure constants of a commutative associative product $\circ$. 

Substituting this last expression of the affinors $(\wa)^i_j$ in \eqref{commutativity}, we obtain the condition: 
\beq\label{commutativity2}
[c^i_{lk}c^l_{jm}-c^i_{lm}c^l_{jk}]X_{\alpha}^k 
X_{\beta}^m=0,
\eeq
which is automatically fulfilled since the product $\circ$ is assumed associative. In particular, for affinors $\wa$ of this form, the condition \eqref{symmetry} is tensorial due to Lemma \ref{lemma1}.

In this case we can 
write \eqref{NLPHT} as
\beq\label{PNLHS2}
P^{ij}=\sum_{\alpha=1}^n(X_{\alpha}\circ u_x)^i
\left(\frac{d}{dx}\right)^{\!-1}\!\!\!(X_{\alpha}\circ 
u_x)^j.
\eeq

We also need the following 
\begin{lemma}\label{lemma3}
Suppose that the vector fields $X_{\alpha}$ are solutions of the following equation:
\beq\label{symmetriesof}
c^i_{jk}\nabla_k X_{\alpha}^l=c^i_{kl}\nabla_j X_{\alpha}^l.
\eeq
Under this assumption,  substituting the $(1,1)$ tensor fields  $(W_{\alpha})^i_j:=c^i_{jk}X^k_{\alpha}$ into \eqref{symmetry} one obtains:$$
(
c^m_{qs}\na_mc^k_{lt}+c^m_{ls}\na_mc^k_{qt}+c^k_{mt}
\na_lc^m_{qs}+c^k_{mt}\na_qc^m_{ls}$$
\beq\label{symmetry2}
-c^m_{qt}\na_mc^k_{ls}-c^m_{lt}\na_mc^k_{qs}-c^k_{ms}
\na_lc^m_{qt}-c^k_{ms}\na_qc^m_{lt}
)X_{\beta}^sX_{\alpha}^t=0.
\eeq
\end{lemma}

\proof
Substituting $(W_{\alpha})^i_j:=c^i_{jk}X^k_{\alpha}$ into \eqref{symmetry}, one needs to show that the terms involving the covariant derivatives of the vector fields $X_{\alpha}$ cancel out. These terms cancel out if and only if the following equality holds: 
$$c^m_{pq}X^p_{\beta}c^k_{ln}\nabla_mX^n_{\alpha}+c^m_{lp}X^p_{\beta}c^k_{qn}\nabla_m X^n_{\alpha}+c^k_{mp}X^p_{\alpha}c^m_{qn}\nabla_l X^n_{\beta}+c^k_{mp}X^p_{\alpha}c^m_{ln}\nabla_qX^n_{\beta}=$$
$$=c^m_{pq}X^p_{\alpha}c^k_{ln}\nabla_mX^n_{\beta}+c^m_{lp}X^p_{\alpha}c^k_{qn}\nabla_m X^n_{\beta}+c^k_{mp}X^p_{\beta}c^m_{qn}\nabla_l X^n_{\alpha}+c^k_{mp}X^p_{\beta}c^m_{ln}\nabla_qX^n_{\alpha}.$$

First we show that $T_1:=X^p_{\beta}c^{m}_{pq}c^k_{ln}\nabla_m X^n_{\alpha}-X^p_{\beta}c^k_{mp}c^m_{qn}\nabla_l X^n_{\alpha}=0$. Indeed, using property \eqref{symmetriesof} applied to the first addendum, we have that $T_1$ is equal to 
$$T_1=X^p_{\beta}(c^m_{pq}c^k_{mn}-c^k_{mp}c^m_{qn} )\nabla_l X^n_{\alpha}$$ 
and this is zero because of associativity of $\circ$. The same procedure works to show that $T_2:=X^p_{\beta}c^m_{lp}c^k_{qn}\nabla_m X^n_{\alpha}-X^p_{\beta}c^k_{mp}c^m_{ln}\nabla_q X^n_{\alpha}$ is also zero. 
Then we have remaining $T_3:=X^p_{\alpha}(c^k_{mp}c^m_{qn}\nabla_l X^n_{\beta}-c^m_{lp}c^k_{qn}\nabla_m X^n_{\beta}).$ Applying this time equality \eqref{symmetriesof} to both addenda of $T_3$ we get
$$T_3=X^p_{\alpha}(c^k_{mp}c^m_{ln}-c^m_{lp}c^k_{mn})\nabla_q X^n_{\beta}=0,$$
again because of the associativity. The same strategy works to show that the remaining term $T_4:=X^p_{\alpha}(c^k_{mp}c^m_{ln}\nabla_qX^n_{\beta}-c^m_{qp}c^k_{ln}\nabla_mX^n_{\beta})$ is also zero. Therefore under the assumption that \eqref{symmetriesof} is satisfied, we have that \eqref{symmetry2} is equivalent to \eqref{symmetry}. 

\begin{flushright}
$\Box$
\end{flushright}

\begin{remark}
In the semisimple case, the compatibility conditions for equations \eqref{symmetriesof} are provided by \eqref{rc+rc+rc}. Condition \eqref{rc+rc+rc} appears also in the non-semisimple case, but it does seem to be sufficient to guarantee the existence of non-trivial solutions of \eqref{symmetriesof}. See \cite{LPR} for details. 

\end{remark}

Finally substituting into 
(\ref{zerocurv}) we obtain the condition
\begin{eqnarray}\label{zerocurv2}
\sum_{\alpha}\epsilon_{\alpha}(c^i_{\,kl}c^j_{\,hm}-c^i_{\,hl}c^j_{\,km})(X_{\alpha})^l(X_{\alpha})^m
=(c^i_{\,kl}c^j_{\,hm}-c^i_{\,hl}c^j_{\,km})g^{lm}=0.
\end{eqnarray}
where $g^{lm}=\sum_{\alpha}\epsilon_{\alpha}(X_{\alpha})^l(X_{\alpha})^m$.

Suppose that $g$ is non-degenerate.
We want to determine sufficient conditions on the metric $g$, on the structure constants $c^i_{jk}$ and on the vector fields $X_{\alpha}$ to ensure that the equations \eqref{symmetry2}, \eqref{commutativity2}, \eqref{zerocurv2} are satisfied, i.e. sufficient conditions ensuring that the bivector \eqref{PNLHS2} is Poisson.  

Let us recall the following definition: 

\begin{definition}\label{deff}
An $F$-manifold is a pair $(M,\circ)$, where $M$ is a manifold and $\circ$ is a commutative associative product 
on vector fields induced by a $(1,2)$ tensor field 
$c$ which satisfies the following condition: 
\beq\label{HertlingManin}
c^m_{tl}(\d_mc^k_{qs})-(\d_m 
c^k_{tl})c^m_{qs}+(\d_qc^m_{tl})c^k_{ms}+(\d_sc^m_{tl})c^k_{mq}-(\d_lc^m_{qs})c^k_{tm}
-(\d_tc^m_{qs})c^k_{lm}=0.
\eeq
\end{definition}
$F$-manifolds were introduced in \cite{HM} as a weak version of Frobenius manifolds. In the original definition
 the product has a unity $e$:
\beq\label{unity}
c^i_{jk}e^k=\delta^i_j
\eeq
However, we prefer to drop this assumption since it does not play any role in this paper.

Some extra structures coming from the theory of integrable systems of hydrodynamic type are
\begin{enumerate}
\item A   symmetric connection $\nabla$
such that 
\beq\label{productcompconnection}
\nabla_l c^i_{jk}=\nabla_j c^i_{lk},
\eeq
and
\beq\label{rc+rc+rc}
R^k_{lmi}c^n_{pk}+R^k_{lip}c^n_{mk}+R^k_{lpm}c^n_{ik}=0,\qquad\forall i,l,m,n,p.
\eeq
\item
A metric $g$ such that $\nabla g=0$ and such that $g$ is invariant with respect to $\circ$,  namely \beq\label{metricinvarproduct}g^{lm}c^j_{hm}=g^{jm}c^l_{hm}.\eeq
\end{enumerate}
An $F$ manifold satisfying conditions \eqref{productcompconnection} and \eqref{rc+rc+rc} is called
 \emph{$F$-manifold with compatible connection} and \emph{Riemannian $F$-manifold} if also the last condition is satisfied.
 A flat Riemannian $F$-manifold with flat unity (see \cite{manin, GLR}) is also called \emph{non-conformal Frobenius manifold}.

Notice that in the definition of $F$-manifold a key condition is given by the so called Hertling-Manin condition (formula \eqref{HertlingManin}), however, in the case of a Riemannian $F$-manifold, this is already implied by the compatibility of $\circ$ with $\nabla$, namely by \eqref{productcompconnection} (see \cite{HeM}). 

We also need the following well-known property: 
\begin{lemma}\label{lemma2}
If the $(1,2)$ tensor field $c$ describes an associative product, then the Hertling-Manin condition \eqref{HertlingManin} is tensorial and is equivalent to 
\beq\label{HMtensorial}
c^m_{tl}(\nabla_mc^k_{qs})-(\nabla_m 
c^k_{tl})c^m_{qs}+(\nabla_qc^m_{tl})c^k_{ms}+(\nabla_sc^m_{tl})c^k_{mq}-(\nabla_lc^m_{qs})c^k_{tm}
-(\nabla_tc^m_{qs})c^k_{lm}=0,
\eeq
where $\nabla$ is an arbitrary symmetric connection.
\end{lemma}
We omit the proof since this is a straightforward computation. We are now ready to prove the main result of this section. 

\begin{theorem}
Let $M$ be an $F$-manifold. Suppose that there exists a connection $\nabla$, not necessarily flat, satisfying condition \eqref{productcompconnection}. If a contravariant metric $g$ admits the following quadratic expansion
$$g=\sum_{\alpha}\epsilon_{\alpha}X_{\alpha}\otimes X_{\alpha},$$
 it is invariant with respect to the product $\circ$, and the vector fields $X_{\alpha}$ are related to $\nabla$ and to $\circ$ via \eqref{symmetriesof}, then the non-local operator \eqref{PNLHS2} is Hamiltonian.
\end{theorem}

\n
{\bf Proof}. 
Since $c$ describes an associative product we can use the Hertling-Manin condition in tensorial form. Also since condition \eqref{commutativity} is automatically satisfied in this case, again because of the associativity, we can use condition \eqref{symmetry} in tensorial form.  
The tensorial Hertling-Manin condition gives
$$HM^k_{lqst}=c^m_{tl}(\nabla_mc^k_{qs})-(\nabla_m 
c^k_{tl})c^m_{qs}+(\nabla_qc^m_{tl})c^k_{ms}+(\nabla_sc^m_{tl})c^k_{mq}-(\nabla_lc^m_{qs})c^k_{tm}
-(\nabla_tc^m_{qs})c^k_{lm}=0,$$
where $\nabla$ is specified as in the statement of the theorem. 
Exploiting the associativity of the product and the symmetry of the tensor $\nabla c$ we have 
\begin{eqnarray*}
&&-(HM^k_{lqst}+HM^k_{qlst})=\\
&=&c^m_{qs}\,\nabla_mc^k_{lt}+c^m_{ls}\,\nabla_mc^k_{qt}-c^k_{ms}\,
\nabla_lc^m_{qt}-c^k_{ms}\,\nabla_qc^m_{lt}+\\
&&-c^m_{qt}\,\nabla_mc^k_{ls}-c^m_{lt}\,\nabla_mc^k_{qs}+c^k_{mt}\,
\nabla_lc^m_{qs}+c^k_{mt}\,\nabla_qc^m_{ls}+\\
&&-c^k_{mq}\,\nabla_s c^m_{tl}+c^k_{lm}\,\nabla_t c^m_{qs}\,
+c^k_{qm}\,\nabla_t c^m_{ls}-c^k_{ml}\, \nabla_s c^m_{tq}\\
&=&c^m_{qs}\,\nabla_mc^k_{lt}+c^m_{ls}\,\nabla_mc^k_{qt}+c^k_{mt}
\,\nabla_lc^m_{qs}+c^k_{mt}\,\nabla_qc^m_{ls}+\\
&&+(-c^m_{qt}\,\nabla_mc^k_{ls}-c^m_{lt}\,\nabla_mc^k_{qs}-c^k_{ms}
\,\nabla_lc^m_{qt}-c^k_{ms}\,\nabla_qc^m_{lt})\\
&=&0.
\end{eqnarray*}
This proves that condition \eqref{symmetry2} is satisfied. By Lemma \ref{lemma3} we know that if the  vector fields $X_{\alpha}$, the connection $\nabla$ and $\circ$ are related via \eqref{symmetriesof}, then condition \eqref{symmetry2} is equivalent to \eqref{symmetry} for the $(1,1)$-tensors $W_{\alpha}$.

Condition \eqref{commutativity2} follows immediately from associativity, while the last condition \eqref{zerocurv2}
 reduces to associativity using the invariance of $g$ with respect to the product.

\begin{flushright}
$\Box$
\end{flushright}
Notice that in the previous Theorem there are {\it no assumptions} on the flatness of the connection $\nabla$ or on the semisimplicity of $\circ$. Due to Remark 2.3 it is natural to assume that $M$ is a Riemannian $F$-manifold.
As a particular case, assuming flatness of $\nabla$, we get the following result obtained in a different way by Mokhov \cite{mokhov}:

\begin{corollary}
Let $(M,\eta,\circ)$ a flat Riemannian $F$-manifold, then the non-local operator, defined in flat coordinates 
$(u^1,\dots,u^n)$ by
\beq\label{PNLHS} 
P^{ij}=\eta^{lm}(u_x\circ)^i_l
\left(\frac{d}{dx}\right)^{\!-1}\!\!\!(u_x\circ)^j_m
\eeq
is a purely non-local Hamiltonian operator.
\end{corollary}

%

\section{Bi-Hamiltonian interpretation of the recursion relations for the principal hierarchy}

Given a flat $F$-manifold one can define an integrable hierarchy of PDEs of the form
$$u^i_{t_{(p,\alpha)}}=c^i_{jk}X^k_{(p,\alpha)}(u)u^j_x,$$
where $p=1,\dots,{\rm dim}(F)$ and $\alpha=0,1,2,\dots$. The vector fields $X^k_{(p,\alpha)}$ defining the hierarchy
 are defined in the following way:
\begin{itemize}
\item $X^k_{(p,0)},p=1,\dots,n$ is a basis of parallel vector fields.
\item starting from $X_{(p,0)}$  one defines $X_{(p,\alpha)}$ recursively as
$$\nabla_j X^i_{(p,\alpha)}=c^i_{jk}X^k_{(p,\alpha-1)}.$$
\end{itemize}
The flows of the principal hierachy can be also written as
$$u^i_{t_{(p,\alpha)}}=\left(g^{il}\d_x+\Gamma^{il}_{j}u^j_x\right)(\omega_{(p,\alpha+1)})_l,$$
where $(\omega_{(p, \alpha+1)})_l=g_{il}X^i_{(p, \alpha+1)}$.
In the Riemannian case, the metric $g$ is invariant with respect to the product: in this case 
$\omega_{(p, \alpha)}=\delta H_{(p, \alpha)}$ ($\alpha\ge 1$).
This means that, in the Riemannian case the flows of the principal hierarchy are Hamiltonian with respect to Poisson brackets
 defined by the local Hamiltonian operator (introduced in \cite{dn84}).
\beq\label{localPoisson}
P^{ij}:=g^{ij}\partial_x+\Gamma^{ij}_ku^k_x
\eeq
Moreover the Hamiltonian functionals are in involution with respect to the associated Poisson bracket:
$$\{H_{(p, \alpha)},H_{(q, \beta)}\}=\int_{S_1}\frac{\delta H_{(p, \alpha)}}{\delta u^i}
\left(g^{il}\d_x+\Gamma^{il}_{j}u^j_x\right)\frac{\delta H_{(q, \beta)}}{\delta u^l}\,dx=0.$$
In the Riemannian case, we have also the non-local Hamiltonian operator \eqref{PNLHS2}. It turns out that the operators \eqref{localPoisson} and \eqref{PNLHS2} are compatible (see \cite{mokhov, GLR}). It is thus natural to ask how the flows obtained via bi-Hamiltonian recursions are related to the flows of the principal hierarchy.\footnote{In the non-Riemannian case an interpretation of the recurrence relations in terms of bi-differential calculus was given in \cite{AL2012}.} Actually it happens that the recurrence using the operators $P$ and $Q$ splits the recurrence of the principal hierarchy into two chains,
one involving the even indices $\alpha$ and one involving the odd ones.

\begin{theorem}
Let $h_{[p,\alpha]}$ be the Hamiltonian densities of the principal hierarchy. Then
\beq
P\delta H_{[p,\alpha+1]}=Q\delta H_{[p,\alpha-1]}
\eeq
\end{theorem}

\n
\emph{Proof}. 
To simplify computations we work in flat coordinates
  $(v^1,\dots,v^n)$. In these coordinates we have $g^{ij}=\eta^{ij}={\rm constant}$ and $P$ and $Q$ have the form
\begin{eqnarray*}
P^{ij}&=&\eta^{ij}\partial_x\\
Q^{ij}&=&\eta^{pq}(v_x \circ)^i_p \partial_x^{-1}(v_x\circ)^j_q
\end{eqnarray*}

By definition, $h_{[p,-1]}:=\eta_{sl}v^l$. The recursive relations of the principal hierarchy
written in terms of the Hamiltonian densities read
\beq\label{rrph}
\partial_i \partial_j h_{[\alpha, s]}=c^l_{ij}\partial_l h_{[\alpha-1,s]}.
\eeq
Using these relation and the invariance of the metric w.r.t. the product we obtain
\begin{eqnarray*}
[Q\delta H_{[s,\alpha-1]}]^i&=&
\eta^{kp}(v_x\circ)^i_k \partial_x^{-1}(v_x\circ)^j_p\partial_j h_{[s,\alpha-1]}
=\eta^{kp}c^i_{lk}v^l_x\partial_x^{-1}c^j_{pq}v^q_x\partial_j h_{[s,\alpha-1]}= \\
&&\eta^{kp}c^i_{lk}v^l_x\partial_x^{-1}\partial_p \partial_q h_{[s,\alpha]} v^q_x=
\eta^{kp}c^i_{lk}v^l_x\partial_x^{-1}\partial_x \partial_p h_{[s,\alpha]}=\\
&&=\eta^{kp}c^i_{lk}v^l_x\partial_p h_{[s,\alpha]}=\eta^{kp}c^i_{lk}v^l_x\partial_p h_{[s,\alpha]}=
\eta^{ip}\partial_l \partial_p h_{[\alpha+1,s]}v^l_x=\\
&=&[P\delta H_{[s,\alpha+1]}]^i.
\end{eqnarray*}

\begin{flushright}
$\Box$
\end{flushright}

Notice that in the recurrence relations \eqref{rrph} of the principal hierarchy, the Hamiltonian densities $ h$
 are not uniquely defined since the partial derivatives $\d_j h$ are defined up to constants. This  is the same ambiguity
 originated  by the presence of the non-local operator $\d_x^{-1}$ in the bi-Hamiltonian recursion relations.

\section{Kohno connections and $\vee$-systems}
Kohno connections appear in the analysis of fundamental groups of hyperplane complements. Since in our analysis we won't need the 
construction in full generality (see for instance \cite{Lo}, \cite{Toledano} for the general case), we content ourselves to recall just a particular case. 

A hyperplane complement is defined by the following data. After having selected a finite dimensional vector space $V$ (either real or complex) and a finite 
collection of distinct linear hyperplanes $\mathcal{H}:=\{H\}_{H\in \mathcal{H}}$, we set $W:=V\setminus \mathcal{H}$. Moreover, for each hyperplane in $\mathcal
{H}$, we choose matrices $\rho_H\in \mathrm{End(V)}$. 
 
Starting from these data, one constructs the following connection on the trivial fiber bundle $W\times V$ over $W$, which can be
 identified with its tangent bundle $TW$: 
\beq\label{dunkl111}\tilde \nabla=\nabla-\lambda \sum_{H\in \mathcal{H}} \frac{d\alpha_H}{\alpha_H} \rho_H,\eeq
where $\nabla$ is a flat connection on $W$, and where $\alpha_H \in V^{*}$ is a linear form defining the hyperplane $H$, one for each $H \in \mathcal{H}$. In 
particular, the form $\omega_H:=\frac{d\alpha_H}{\alpha_H}$ is acting as follows on a tangent vector $X\in T_p W$: 
$$\omega_H(X):=\frac{\alpha_H(X_p)}{\alpha_H(p)},$$
where $p\in W$ is identified with a vector. 

The following useful criterion by Kohno (see \cite{Ko}, \cite{Toledano}, \cite{Lo}) provides necessary and sufficient conditions for the connection $\tilde \nabla$ to be 
flat:
\begin{proposition}
The connection \eqref{dunkl111} is flat for any $\lambda$ if and only if, for any collection of linear forms $\{\alpha_H\}_{H\in \mathcal{H}'}$ which is maximal for the property that their span in $V^
{*}$ is two-dimensional, one has 
$$[\sum_{H\in \mathcal{H}'} \rho_H, \rho_K]=0,$$
for each $\rho_K$ with $K\in \mathcal{H}'.$
\end{proposition}
To relate the flatness of a Kohno connection with the condition characterizing  $\vee$-systems, we will need a more specialized construction, which we recall 
following in part \cite{Lo}. 

Suppose $V$ is also equipped with a non-degenerate symmetric bilinear form $g$ with the property that for each $H\in \mathcal{H}$ the vector space $V$ is 
decomposed as a direct sum $V=H\oplus H^{\perp}$, where $H^{\perp}$ is the $g$-orthogonal complement of $H$ (this means that $g$ restricted to $H\times H$ is also non-degenerate for each $H\in \mathcal{H}$). 
For each $H\in \mathcal{H}$, we define the endomorphism $\rho_H$ as $\rho_H:=\alpha_H \otimes \check{\alpha}_H$, where $\alpha_H\in V^*$ is a linear form 
with $\mathrm{Ker}(\alpha_H)=H$ as above and $\check{\alpha}_H$ is a vector in $H^{\perp}$ such that $\alpha_H=g(\cdot, \check{\alpha}_H).$

It is immediate to see that the $\rho_H$ constructed in this way are self-adjoint with respect to $g$, since $g(v, \rho_H(w))=\alpha_H(v)\alpha_H(w)=g(\rho_H(v), w)$.

Moreover, since the endomorphisms $\rho_H$ have a special form in this case, it is possible to define a {\it commutative} product on the tangent bundle of $W$ as follows (see \cite{Lo}), where 
$X_p, Y_p\in T_pW$:
\begin{equation}\label{product1}
X_p\cdot Y_p:=\sum_{H\in \mathcal{H}} \omega_H(X_p) \rho_H(Y_p)=\sum_{H\in \mathcal{H}}(\alpha_H(p))^{-1}\alpha_H(X_p)\alpha_H(Y_p) \check{\alpha}_H.
\end{equation}
It is clear that this product is automatically commutative and if $\sum_{H\in \mathcal{H}} \rho_H=\mu \mathrm{Id}$, for some constant $\mu\neq 0$, then the 
rescaled Euler vector field $\tilde E_p:=\frac{1}{\mu} E_p=(p,\frac{1}{\mu} p)$ is the identity for the product \eqref{product1}: 
$$X_p\cdot \tilde E_p=\frac{1}{\mu}\sum_{H\in \mathcal{H}}\alpha_H(X_p) \alpha_H(p)(\alpha_H(p))^{-1}\check{\alpha}_H=\frac{1}{\mu}\sum_{H\in \mathcal{H}} \rho_H(X_p)=X_p.$$
(Here, the notation $(p, \frac{1}{\mu}p)$ indicate the fact that the vector $\frac{1}{\mu}p$ has to be thought as belonging to the vector space $T_pW$; so the first $p$ is just the base point of the vector, while $\frac{1}{\mu}p$ is the vector itself.)
Using the product \eqref{product1}, it is possible to construct a deformed connection 
\begin{equation}\label{deformedconnection1}
\tilde \nabla_X Y:=\nabla_X Y+\lambda X\cdot Y
\end{equation}
where $\nabla$ is the flat Levi-Civita connection associated to $g$. 

Kohno's criterion can be applied to this special case and leads to the following definition (see also \cite{Lo}): 
\begin{definition}\label{dunklproperty}
The data $(V, g, \{\rho_H\}_{H\in \mathcal{H}})$ have the {\em Kohno property} if for every linear subspace $L\subset V$
of codimension two obtained as an 
intersection of members of $\mathcal{H}$, the sum 
$\sum_{H\in \mathcal{H},\, L\subset H}\rho_H$ commutes with each of its terms, namely 
\beq\label{dunkl1}
[\sum_{H\in \mathcal{H},\, L\subset H} \rho_H, \rho_K]=0,
\eeq
for each $\rho_K$ appearing as a term in the sum $\sum_{H\in \mathcal{H},\, L\subset H} \rho_H$.
\end{definition}
Clearly, due to the pevious discussion, if the data $(V, g, \{\rho_H\}_{H\in \mathcal{H}})$ have the Kohno property 
the deformed connection \eqref{deformedconnection1} $\tilde \nabla$ is flat for every $\lambda$.
 This means that
\begin{eqnarray*}
&&R^i_{mjl}+\left[\nabla_m c^i_{jl}-\nabla_j c^i_{ml}\right]\lambda+\left[c^i_{jk}c^k_{ml}-c^i_{mk}c^k_{jl}\right]\lambda^2=0,\qquad\forall\lambda.
\end{eqnarray*}
Since $\lambda$ is arbitrary, the product \eqref{product1} is associative and $\nabla_m c^i_{jl}=\nabla_j c^i_{ml}$. 
The last step providing the Frobenius potential is related to the invariance of the metric $g$ w.r.t. the product  \eqref{product1}.
 In this case, the Frobenius potential has the form (see
 for instance \cite{Lo}, \cite{FV1}): 
\beq\label{Fpotential}
F(p):=\frac{1}{2}\sum_{H\in \mathcal{H}}(\alpha_H(p))^2\log{\alpha_H(p)}.\eeq
We are going to show that the conditions of $\vee$-system are exactly the conditions for which the data $(V, g, \{\rho_H\}_{H\in \mathcal{H}})$ have the Kohno 
property, under some very mild assumptions. 

$\vee$-systems were introduced by A. Veselov in \cite{Ve} to construct new solutions of generalized WDVV equations, starting from a special set of covectors. 
The $\vee$-conditions are precisely the conditions that guarantee that a particular function associated to special set of covectors satisfies the WDVV equations. It is known that the $\vee$-conditions are fulfilled for all root systems and for their 
special deformations discovered in the Calogero-Moser systems (see \cite{CFV}). $\vee$-systems have been extensively studied by Veselov and his collaborators (see for instance \cite{FV1}, \cite{FV2}, \cite{CV}).

Now we recall the notion of $\vee$-systems (see \cite{Ve}).
Let $\mathcal{V}$ be a finite set of non-collinear covectors $\alpha\in V^*$. The condition that the covectors are not collinear means that the set of associated 
hyperplanes $\mathcal{H}$, where $H=\mathrm{Ker}(\alpha)$ is made of distinct hyperplanes. We assume that the collection of covectors $\mathcal{V}$ span $V^*
$ so that the symmetric  bilinear form defined by $G^{\mathcal{V}}:=\sum_{\alpha\in  \mathcal{V}} \alpha\otimes \alpha$
is non-degenerate. In particular, the non-degeneracy of $G^{\mathcal{V}}$ is equivalent to ask that the map $\phi_{\mathcal{V}}: V\rightarrow V^*$ defined by the 
formula 
$$(\phi_{\mathcal{V}}(u))(v):=G^{\mathcal{V}}(u,v), \quad u, v\in V,$$
is invertible. In this context, it is possible to define the vector $\check \alpha\in V$ as
\begin{equation}\label{checkalpha}
\check{\alpha}:=\phi^{-1}_{\mathcal{V}}(\alpha), \quad \alpha\in V^*, 
\end{equation}
or, which is equivalent, as the unique vector in $V$ such that 
\begin{equation}\label{checkalpha2}\alpha=G^{\mathcal{V}}(\cdot, \check{\alpha}).\end{equation}
Observe also that the linear map $\sum_{\beta \in \mathcal{V}} \check{\beta}\otimes \beta: V\rightarrow V$ is just the identity map, so that 
$$\alpha(v)=\sum_{\beta \in \mathcal{V}}\alpha(\check{\beta})\beta(v).$$
\begin{definition}(See \cite{Ve})\label{veecondition}
We say that the spanning set $\mathcal{V}:=\{\alpha\}_{\alpha\in \mathcal{V}}\subset V^*$ satisfies the $\vee$-conditions or it is a $\vee$-system if for each two-dimensional plane $
\Pi\subset V^*$ we have
\begin{equation}\label{veeequation}
\sum_{\beta\in \Pi\cap \mathcal{V}}\beta(\check{\alpha})\check{\beta}=\lambda \check{\alpha},
\end{equation}
for each $\alpha \in \Pi\cap \mathcal{V}$ and for some $\lambda$, which may depend on $\Pi$ and $\alpha$. 
\end{definition}

\begin{remark}\label{remarkimportant}
(See \cite{Ve}) Let us remark few points about the definition \ref{veecondition} and the condition \eqref{veeequation}. 

If the plane $\Pi\subset V^*$ contains at most one covector $\beta\in \mathcal{V}$, then condition \eqref{veeequation} is automatically satisfied for that plane. 

If the plane $\Pi\subset V^*$ contains at least three covectors in $\mathcal{V}$, then it is immediate to see that the constant $\lambda$ in \eqref{veeequation} has 
to be the same for all $\alpha \in \Pi\cap \mathcal{V}$, for the given $\Pi$, and therefore condition \eqref{veeequation} simply means that $\sum_{\beta\in \Pi\cap 
\mathcal{V}} \beta\otimes \check{\beta}=\lambda \mathrm{Id}$ on the plane $\check{\Pi}\subset V$. 

The only case in which the constant $\lambda$ in \eqref{veeequation} might depend on $\alpha$ for a given plane $\Pi$ is when $\Pi$ contains exactly two 
covectors from $\mathcal{V}$, $\beta_1$ and $\beta_2$. In this case, the following two equalities must be fulfilled: 
$$\beta_1(\check{\beta}_1)\check{\beta}_1+\beta_2(\check{\beta}_1)\check{\beta}_2=\lambda_{\beta_1}\check{\beta}_1,$$
$$\beta_1(\check{\beta}_2)\check{\beta}_1+\beta_2(\check{\beta}_2)\check{\beta}_2=\lambda_{\beta_2}\check{\beta}_2.$$
These are equivalent to $G^{\mathcal{V}}(\check{\beta}_1, \check{\beta}_2)=0$, since for instance, the first relation above gives $\beta_2(\check{\beta}_1)=0$ and analogously for the second. Moreover, in this case $\lambda_{\beta_1}=\beta_1
(\check{\beta}_1)$ and $\lambda_{\beta_2}=\beta_2(\check{\beta}_2).$
\end{remark}
To a given finite spanning set $\mathcal{V}:=\{\alpha\}_{\alpha\in \mathcal{V}}\subset V^*$, in which the covectors $\alpha$ are non-collinear, we can associate a 
finite collection $\mathcal{H}$ of distinct hyperplanes identified by the kernels of the covectors $\alpha$ (so that $\alpha=\alpha_H$ for the corresponding 
hyperplane), a non-degenerate positive definite bilinear form $g:=G^{\mathcal{V}}$ and a collection of rank one endomorphisms $\{\rho_H:=\alpha_H\otimes \check
{\alpha}_H\}_{H\in \mathcal{H}}$. The following theorem clarifies the relation between $\vee$-systems and data $(V, g, \{\rho_H\}_{H\in \mathcal{H}})$ having the 
Kohno property. Its content is not surprising, since one can indirectly prove it using the equivalence with WDVV equations (see the Remark at the end of Section I of \cite{Ve2}),
however the proof we provide here is completely elementary and direct, being based on Linear Algebra. A similar result has been obtained
 independently by M.V. Feigin and A.P. Veselov and will appear in a forthcoming paper \cite{FV}.
 
\begin{theorem}\label{veeKohno}
Assume that in the finite spanning set $\mathcal{V}:=\{\alpha\}_{\alpha\in \mathcal{V}}\subset V^*$ the covectors $\alpha$ are non-collinear and associate to $
\mathcal{V}$ a finite collection $\mathcal{H}$ of distinct hyperplanes $H\subset V$.
Then the data $(V, g:=G^{\mathcal{V}}, \{\rho_H:=\alpha_H\otimes \check{\alpha}_H\}_{H\in \mathcal{H}})$ have the Kohno
property if and only if the spanning set $\mathcal{V}$ is a $\vee$-system. 
\end{theorem}

\proof
First, we observe the following. For each $2$-plane $\Pi\subset V^*$, consider the codimension two annihilator $\mathrm{Ann}(\Pi)\subset V^{**}$. Using the 
canonical isomorphism of $V^{**}$ with $V$, we can identify $\mathrm{Ann}(\Pi)$ with a codimension two subspace in $V$, call it $L_{\Pi}$. Observe also that for $
\alpha\in \mathcal{V}$ we have:  $\alpha\in \Pi$, for a given $\Pi$ if and only if the hyperplane $H_{\alpha}$ corresponding to $\alpha$ ($H_{\alpha}:=\mathrm{Ker}
(\alpha)$) contains $L_{\Pi}$. Moreover, if we consider the $G^{\mathcal{V}}$-orthogonal of $L_{\Pi}$ in $V$, we have that $(L_{\Pi})^{\perp}=\check{\Pi}$. 
Indeed, take any $w\in L_{\Pi}$ and consider $g(w, \check{\alpha})=G^{\mathcal{V}}(w, \check{\alpha})=\alpha(w)=0$, since $w\in L_{\Pi}=\mathrm{Ann}(\Pi)$. 

Now assume that the spanning set of non-collinear covectors $\mathcal{V}$ is a $\vee$-system. To each $\Pi\subset V^*$, we consider the corresponding $L_{\Pi}
$. If $\Pi$ contains no covector $\alpha\in \mathcal{V}$, this means that $L_{\Pi}$ is not contained in any hyperplane $H_{\alpha}$, so no conditions need to be 
checked. 

If $\Pi$ contains exactly one covector $\alpha\in \mathcal{V}$, this means that $L_{\Pi}$ is contained in exactly one hyperplane $H_{\alpha}$. In this case the
 Kohno condition is trivially satisfied, since 
$[\rho_{H_{\alpha}}, \rho_{H_{\alpha}}]=0.$

If $\Pi$ contains three or more covectors $\alpha\in \mathcal{V}$, then by Remark \ref{remarkimportant}, $\sum_{\alpha \in \mathcal{V}\cap \Pi} \alpha\otimes \check{\alpha}
=\lambda \mathrm{Id}$ on $\check{\Pi}$. Now  $\sum_{\alpha \in \mathcal{V}\cap \Pi} \alpha\otimes \check{\alpha}=\sum_{L_{\Pi}\subset H_{\alpha}} \rho_{H_
{\alpha}}$. Moreover,  $V=L_{\Pi}\oplus (L_{\Pi})^{\perp}=L_{\Pi}\oplus \check{\Pi}$ and  ${\rho_{H_{\alpha}}}_{|L_{\Pi}}=0$ for each $H_{\alpha}$ with $L_{\Pi}
\subset H_{\alpha}$, since $\alpha$ annihilates $H_{\alpha}$ by definition and $L_{\Pi}$ is contained in all the hyperplanes $H_{\alpha}$ with $\alpha \in \mathcal
{V}\cap \Pi$. This implies that $\sum_{L_{\Pi}\subset H_{\alpha}} {\rho_{H_{\alpha}}}_{|L_{\Pi}}=0$.  Therefore  the Kohno condition $[\sum_{L_{\Pi}\subset H_
{\alpha}} \rho_{H_{\alpha}}, \rho_{H_{\alpha}}]=0$ is automatically satisfied on the codimension two subspace $L_{\Pi}$ and we need to check it only on $(L_{\Pi})^
{\perp}=\check{\Pi}$, since $V=L_{\Pi}\oplus \check{\Pi}$ and $\rho_{H_{\alpha}}(L_{\Pi})\subset (L_{\Pi})$ and $\rho_{H_{\alpha}}(\check{\Pi})\subset \check{\Pi}
$, so the decomposition is invariant under $\rho_{H_{\alpha}}$.  But on $\check{\Pi}$ we have that $\sum_{L_{\Pi}\subset H_{\alpha}} \rho_{H_{\alpha}}$ is a multiple of the identity, so again the Kohno 
condition is fulfilled.

Finally if $\Pi$ contains exactly two covectors $\alpha_1, \alpha_2\in \mathcal{V}$, by Remark \ref{remarkimportant} we have that they are $g$-orthogonal to each
 other. Therefore, if we write the linear operator $\alpha_1\otimes \check{\alpha}_1+\alpha_2\otimes \check{\alpha}_2$ restricted to $\check{\Pi}$ in the basis $
 (\check{\alpha}_1, \check{\alpha}_2)$, we see that it is diagonal (not necessarily a multiple of the identity). Moreover, each of the operators $\alpha_i\otimes \check
 {\alpha_i}$, $i=1,2$ restricted to $\check{\Pi}$ in the basis $(\check{\alpha}_1, \check{\alpha}_2)$ is also diagonal (with one eigenvalue zero). Therefore in this 
 case the Kohno condition  
$[\rho_{H_{\alpha_1}}+\rho_{H_{\alpha_2}}, \rho_{H_{\alpha_i}}]=0$, $i=1,2$ is satisfied because on $(L_{\Pi})^{\perp}=\check{\Pi}$ all these operators are 
simultaneously diagonal, while on $L_{\Pi}$ it is trivially satisfied since they are all zero. 

This proves that if $\mathcal{V}$ is a $\vee$-system, then the data $
(V, g:=G^{\mathcal{V}}, \{\rho_H:=\alpha_H\otimes \check{\alpha}_H\}_{H\in \mathcal{H}})$ have the Kohno property. 

Viceversa, suppose these data have the Kohno property. We show that the covectors $\alpha$ satisfy the $\vee$-condition. 

Fix a codimension two subspace $L\subset V$ and consider all hyperplanes $H$ of the form $H_{\alpha}$ containing $L$. 
Kohno condition means $[\sum_{L\subset H_{\alpha}} \rho_{H_{\alpha}}, \rho_{H_{\alpha}}]=0$. Observe that this is automatically fulfilled on $L$, since for each $w
\in L$, $\rho_{H_{\alpha}}(w)=0$ if $L\subset H_{\alpha}$. So the Kohno condition is significant only on the $G^{\mathcal{V}}$-orthogonal complement $L^{\perp}$ 
of $L$ in $V$. As before, to each $L$ of codimension two we can associate the $2$-plane $\mathrm{Ann}(L)\subset V^*$, call it $\Pi_L$. To each hyperplane $H_
{\alpha}$ containing $L$ there is a corresponding covector $\alpha \in \Pi_L$. 

If for the fixed $L$ there is at most one hyperplane of the form  $H_{\alpha}$ containing it, then Kohno condition is trivially satisfied, and also the $\vee$-condition 
since in this case it simply reads $\alpha(\check{\alpha})\check{\alpha}=\lambda\check{\alpha}$, which is true with $\lambda= \alpha(\check{\alpha}).$

Suppose that for the fixed $L$ there are exactly two hyperplanes containing it, namely $H_{\alpha_1}$ and $H_{\alpha_2}$. This means $\alpha_1, \alpha_2\in 
\Pi_L\subset V^*$ and the nontrivial part of the Kohno condition reads $[\rho_{H_{\alpha_1}}+ \rho_{H_{\alpha_2}}, \rho_{H_{\alpha_i}}]_{|L^{\perp}}=0$ for $i=1,2$. 
This is the same as 

$[\rho_{H_{\alpha_1}},  \rho_{H_{\alpha_2}}]_{|\check{\Pi}_L}=0.$ Since $\check{\Pi}_L$ is invariant for $\rho_{H_{\alpha_i}}$, $i=1,2$, and $\{\check{\alpha}_1, 
\check{\alpha}_2\}$ are a basis for $\check{\Pi}_L$ it is immediate to see that  $[\rho_{H_{\alpha_1}},  \rho_{H_{\alpha_2}}](\check{\alpha}_1)=0$ and $[\rho_{H_
{\alpha_1}},  \rho_{H_{\alpha_2}}](\check{\alpha}_2)=0$ if and only if $G^{\mathcal{V}}(\check{\alpha}_1, \check{\alpha}_2)=0$. But this means that the operator 
$\alpha_1\otimes \check{\alpha}_1+\alpha_2\otimes\check{\alpha}_2$ restricted to $\check{\Pi}_L$ is diagonal with possibly distinct eigenvalues. This is exactly the
 $\vee$-condition in this case. 

Finally, suppose that for fixed $L$, there are three or more hyperplanes of the form $H_{\alpha}$ containing it. This means that $\Pi_L$ contains three or more of 
the covectors $\alpha\in \mathcal{V}$. Call them $\{\alpha_1, \dots, \alpha_n\}$ with $n\geq 3$. Now Kohno condition gives
$$\left[\sum_{i=1}^n \alpha_i\otimes \check{\alpha}_i, \alpha_j\otimes \check{\alpha}_j\right]_{|\check{\Pi}_L}=0, \quad j=1, \dots, n.$$
Developing $\left[\sum_{i=1}^n \alpha_i\otimes \check{\alpha}_i, \alpha_j\otimes \check{\alpha}_j\right](\check{\alpha}_j)=0$ in a suitable way we get immediately 
$$\alpha_j(\check{\alpha}_j)\left(\left( \sum_{i=1}^n\alpha_i\otimes \check{\alpha}_i \right)(\check{\alpha}_j)\right)-\alpha_j\left( \left( \sum_{i=1}^n \alpha_i\otimes 
\check{\alpha}_i\right)(\check{\alpha}_j)\right)\check{\alpha}_j=0,$$
which means that 
$$\left(\sum_{i=1}^n \alpha_i\otimes \check{\alpha}_i\right)(\check{\alpha}_j)=\frac{\alpha_j\left(\left(\sum_{i=1}^n \alpha_i\otimes \check{\alpha}_i\right)(\check
{\alpha}_j) \right)}{\alpha_j(\check{\alpha}_j)} \check{\alpha}_j.$$
Equivalently each $\check{\alpha}_j$ is an eigenvector of $\sum_{i=1}^n \alpha_i\otimes \check{\alpha}_i$. Therefore the linear operator  $\sum_{i=1}^n \alpha_i
\otimes \check{\alpha}_i$ on the two dimensional space $\check{\Pi}_L$ has more then two eigenvectors, since $n\geq 3$ and each $\check{\alpha}_j\in \check{\Pi}
_L$, and this is possible iff $\sum_{i=1}^n \alpha_i\otimes \check{\alpha}_i$ is a scalar multiple of the identity. This is exactly the $\vee$-condition in this case.  

\begin{flushright}
$\Box$
\end{flushright}

\subsection{Two examples}

Here we present two examples of $\vee$-systems that depends on parameters and that become singular for certain values of the parameters. Nevertheless we show that even in the singular case, we can obtain a potential which is a sort of regularized potential. 
We start with an example corresponding to the exceptional generalized  root system $D(2,1,\lambda)$, taken from \cite{FV1}. The $\vee$-system $\mathcal{V}$ in this case consists of the following seven covectors $\alpha$ in $\mathbb{C}^3$: 
$$e_1\pm e_2\pm e_3, \, \sqrt{2(t+s-1)}e_1, \sqrt{\frac{2(s-t+1)}{t}}e_2, \sqrt{\frac{2(t-s+1)}{s}}e_3,$$
where $t,s$ are parameters related to the projective parameter $\lambda=[\lambda_1:\lambda_2:\lambda_3]$ as affine coordinates in the chart $\lambda_1\neq 0.$ 
The symmetric  bilinear form defined by $G^{\mathcal{V}}:=\sum_{\alpha\in  \mathcal{V}} \alpha\otimes \alpha$ becomes 
$$G^{\mathcal{V}}:=\left[\begin{array}{ccc} 
2(t+s+1) & 0 & 0\\
0 & \frac{2(t+s+1)}{t} & 0\\
0 & 0 & \frac{2(t+s+1)}{s}
  \end{array} \right]$$
which vanishes if $t+s+1=0$.  For $t+s+1\ne 0$ the (non conformal) Frobenius structure $(\eta,\circ,e)$ is defined by the following data:
\begin{itemize}
\item $\eta(s,t)= G^{\mathcal{V}}$.
\item the structure constants of the product are defined by the formula
$$c^i_{jk}(s,t)=(\eta^{-1})^{il}\d_l\d_j\d_k F,$$
where the potential is given by the standard formula \eqref{Fpotential}
\beq\label{FP}
F(s,t)=\f{1}{2}\sum_{\alpha\in\mathcal{V}}\alpha(p)^2\ln\alpha(p)
\eeq
\item If we call $H_{\alpha}$ the hyperplane defined by $\alpha$ it is easy to check that $\sum_{\alpha\in \mathcal{V}} \rho_{H_{\alpha}}=\mathrm{Id}$. Thus the unit vector field at the point $p$ is given by the formula $e_p=(p,p)$.
\end{itemize}
The case  $t+s+1=0$ is more delicate. Indeed the metric vanishes and the structure constants $c_{ijk}$ are  rational functions depending on parameters $s,t$ and they all contain a factor $(s+t+1)$ at the denominator,
 signaling the presence of a singularity for $s+t+1=0$. We can regularize the structure constants multiplying by $(s+t+1)$ and evaluating at  $t=-s-1$.
 Symmetrically we can regularize the metric dividing by $(t+s+1)$  and evaluating at  $t=-s-1$. The final result is the following:
\begin{itemize}
\item $\eta$ is given by the regularized metric
$$G^{\mathcal{V}}_{reg}=\left[\begin{array}{ccc} 
2 & 0 & 0\\
0 & \frac{2}{t} & 0\\
0 & 0 & -\frac{2}{1+t}
  \end{array} \right]$$
\item the regularized structure constants of the product are defined by the formula
$$(c_{reg})^i_{jk}=(\eta^{-1})^{il}\d_l\d_j\d_k F_{reg},$$
where the regularized potential is still given by the standard formula \eqref{Fpotential}
$$F_{reg}=\f{1}{2}\sum_{\alpha\in\mathcal{V}}\alpha(p)^2\ln\alpha(p).$$

This can be easily explained observing that 
\begin{eqnarray*}
\d_i\d_j\d_k F_{reg}&=&\lim_{s+t+1\to 0}\d_i\d_j\d_k F(s,t)=\\
&&\lim_{s+t+1\to 0}\eta_{il}(s,t)c^l_{jk}(s,t)=\\
&&\lim_{s+t+1\to 0}\frac{\eta_{il}(s,t)}{s+t+1}(s+t+1)c^l_{jk}(s,t)=(G^{\mathcal{V}}_{reg})_{il}(c_{reg})^l_{jk}.
\end{eqnarray*}
\item In this case is easy to check that $\sum_{\alpha\in \mathcal{V}} \rho_{H_{\alpha}}=0$ and thus the product loses the unity. 
\end{itemize}

A second example always from \cite{FV1} corresponds to the generalized root system $G(1,2)$. In this case we have: 
\begin{eqnarray*}
&&\sqrt{2t+1}e_1,\,\sqrt{2t+1}e_2,\,\sqrt{2t+1}(e_1+e_2),\,\sqrt{\f{2t-1}{3}}(e_1-e_2),\,\sqrt{\f{2t-1}{3}}(2e_1+e_2),\\
&&\sqrt{\f{2t-1}{3}}(e_1+2e_2), \sqrt{\frac{3}{t}}e_3, \, e_1\pm e_3,\,e_2\pm e_3,\,e_1+e_2\pm e_3.
\end{eqnarray*}
The symmetric  bilinear form defined by $G^{\mathcal{V}}:=\sum_{\alpha\in  \mathcal{V}} \alpha\otimes \alpha$ becomes 
$$G^{\mathcal{V}}:=\left[\begin{array}{ccc} 
4(2t+1) & 2(2t+1) & 0\\
2(2t+1) & 4(2t+1) & 0\\
0 & 0 & \frac{3(2t+1)}{t}
  \end{array} \right]$$
which vanishes at $t=-1/2$. For $t\ne -1/2$ we have a non conformal Frobenius manifold defined as above. For $t=-1/2$ the first $3$ covectors vanish,
 the regularized metric is given by
$$G^{\mathcal{V}}_{reg}=\left[\begin{array}{ccc} 
1 & \f{1}{2} & 0\\
\f{1}{2} & 1 & 0\\
0 & 0 & -\frac{3}{2}
  \end{array} \right],$$
the regularized structure constants are
$$c^i_{jk}=((G^{\mathcal{V}}_{reg})^{-1})^l\d_l\d_j\d_k F,$$
and the potential is given by 
$$F=\f{1}{2}\sum_{\alpha\in\mathcal{V}}\alpha(p)^2\ln\alpha(p),$$
where the sum is taken over the non-vanishing  covectors of $\lim_{t\to-\f{1}{2}} \mathcal{V}(t)$. 
Finally, as in the previous example, in the degenerate limit $t\to-1/2$ the unity $e$
 disappears since also in this case $\sum_{\alpha\in \mathcal{V}} \rho_{H_{\alpha}}=0$. 

We have presented these two examples, taking into account also degenerates cases, to see what happens to the associated non-conformal Frobenius structures. 

\section{$\vee$-systems and purely non-local Hamiltonian structures}

In the previous Section we have seen that a finite spanning set $\mathcal{V}\subset V^*$ of non-collinear covectors satisfying the $\vee$-conditions gives rise to a 
commutative associative product on the vector fields on $V$ and that there is an equivalence between $\vee$-systems and corresponding data fulfilling the Kohno 
condition. 
In this section we work out the purely non-local Hamiltonian structure corresponding to this structure. 

\begin{proposition}\label{purelynonlocal}
Let $V$ be a finite dimensional vector space. Then to any $\vee$-system of covectors $\mathcal{V}\subset V^*$, in particular to any root system, there is a 
associated a purely non-local Hamiltonian structure of the following form: 
\beq\label{PNHSvee}
 P=\sum_{\beta, \gamma\in \mathcal{V}}\check{G}(\beta, \gamma) \partial_x \log(\beta(u)) \partial_x^{-1} (\partial_x \log(\gamma(u))) \check{\beta} \otimes \check{\gamma},
\eeq
where $\check{G}(\cdot, \cdot)$ is the co-metric acting on covectors. 
\end{proposition}
\proof

The metric  $\check{G}$ defined by  the quadratic expansion
$$\check{G}(\beta, \gamma)=\sum_{\alpha\in \mathcal{V}}\beta(\check{\alpha}) \gamma
(\check{\alpha}),$$ 
is invariant with respect to the product
$$c^i_{lp}(u)=\sum_{\alpha\in \mathcal{V}}\frac{\alpha_l \, \alpha_p \, \check{\alpha}^i}{\alpha(u)}.$$
Due to the results of previous section the non-local operator
\beq
P^{ij}=\sum_{\alpha\in\mathcal{V}}(\check{\alpha}\circ u_x)^i
\left(\frac{d}{dx}\right)^{\!-1}\!\!\!(\check{\alpha}\circ 
u_x)^j
\eeq
is a purely non-local Hamiltonian operator. Substituting the expression of the constant structures in the formula above for $P^{ij}$ we get
$$P^{ij}=\sum_{\alpha\in \mathcal{V}}\sum_{\beta \in \mathcal{V}}\frac{\beta_l \beta_p\check{\beta}^i}{\beta(u)}\check{\alpha}^lu^p_x\partial_x^{-1}\sum_{\gamma\in \mathcal{V}}\frac{\gamma_m \gamma_n \check{\gamma}^j}{\gamma(u)}\check{\alpha}^m u^n_x=$$
$$\sum_{\alpha\in \mathcal{V}}\sum_{\beta\in \mathcal{V}}\beta(\check{\alpha})\check{\beta}^i\partial_x\left(\log((\beta(u))\right)\partial_x^{-1}\sum_{\gamma\in \mathcal{V}}\partial_x \left(\log(\gamma(u))\right)\gamma(\check{\alpha})\check{\gamma}^j.$$
Summing over $\alpha\in \mathcal{V}$ and using the definition of the co-metric we get immediately \eqref{PNHSvee}.

\begin{flushright}
$\Box$
\end{flushright}

Let us remark that the Proposition \ref{purelynonlocal}  applies not only to non-degenerate $\vee$-systems, but also to the degenerate cases, once the singular metric has been substituted with the regularized one, as we have done in the two examples in Section 4.  

\bigskip

{\bf Acknowledgements} We would like to thank Misha Feigin for fruitful discussions and for having pointed out reference \cite{Ve2} 
and Alexander Veselov for  useful comments. This work has been partially supported by the Italian MIUR Research Project
\emph{Teorie geometriche e analitiche dei sistemi Hamiltoniani in dimensioni finite e infinite}.
The research of AA is partially supported by the Faculty Development Funds of the College of Natural Sciences and Mathematics,
University of Toledo.

%

%

\end{document}